\documentclass[12pt]{article}

\usepackage{bm}
\usepackage{amssymb}

\input epsf
\def\beq{\begin{eqnarray}}
\def\eeq{\end{eqnarray}}
\def\e{{\epsilon}}

\def\({\left(}
\def\){\right)}
\def\Earth{\oplus}

\def\mpl{M_{\rm pl}}

\def\e{\epsilon}

\def\lsim{\mathrel{\rlap{\lower3pt\hbox{\hskip0pt$\sim$}}
     \raise1pt\hbox{$<$}}}         %less than or approx. symbol
\def\gsim{\mathrel{\rlap{\lower4pt\hbox{\hskip1pt$\sim$}}
     \raise1pt\hbox{$>$}}}         %greater than or approx. symbol

\def\lsim{\mathrel{\rlap{\lower3pt\hbox{\hskip0pt$\sim$}}
     \raise1pt\hbox{$<$}}}         %less than or approx. symbol
\def\gsim{\mathrel{\rlap{\lower4pt\hbox{\hskip1pt$\sim$}}
     \raise1pt\hbox{$>$}}}         %greater than or approx. symbol

\newcommand{\be}{\begin{equation}}
\newcommand{\ee}{\end{equation}}
\newcommand{\comment}[1]{}
\usepackage{amsmath}
\usepackage{amsfonts}
\usepackage{verbatim}
\usepackage{graphicx}
\usepackage{subfigure}
\usepackage{amssymb}
\usepackage[T1]{fontenc}

\def\ea{\end{eqnarray}}
\def\ba{\begin{eqnarray}}

\def\beq{\begin{eqnarray}}
\def\eeq{\end{eqnarray}}
\def\e{{\epsilon}}

\def\({\left(}
\def\){\right)}
\def\mn{_{\mu \nu}}
\def\e{\mathrm{e}}

\def\mpl{M_{\rm pl}}

\def\e{\epsilon}

\def\d{\delta}
\def\p{\partial}

\def\L{\mathcal L}

\def\d{\mathrm{d}}
\def\mn{_{\mu \nu}}

\def\({\left(}
\def\){\right)}
\def\e{\epsilon}

\def\ie{{\it i.e. }}

\def\mpl{M_{\rm Pl}}
\def\p{\partial}
\def\La{\Lambda_3}
\def\LE{\Lambda_\Earth}

\renewcommand{\comment}[1]{}

\begin{document}

\begin{center}{\Large \bf Mixed Galileons and Spherically Symmetric Solutions}
\end{center}
\vskip 0.2cm
\centerline{\Large \bf }

\vskip 0.7cm
\centerline{\large L.  Berezhiani$^a$, G. Chkareuli$^b$, C. de Rham$^{c}$,
G.  Gabadadze$^{b}$ and A.J. Tolley$^c$}

\vskip 0.3cm

\centerline{\em $^a$Center for Particle Cosmology, Department of Physics and Astronomy,}
\centerline{\em University of Pennsylvania, Philadelphia, Pennsylvania 19104, USA}

\centerline{\em $^b$Center for Cosmology and Particle Physics,
Department of Physics,}
\centerline{\em New York University, New York,
NY, 10003}

\centerline{\em $^c$Department of Physics, Case Western Reserve University,
Euclid Ave,
Cleveland, OH, 44106}

\vskip 1.9cm

\begin{abstract}

% This is the sequel to our recent paper arXiv:1302.0549; which studied the Vainshtein mechanism in the decoupling limit of massive gravity.
It was previously found that in a certain parameter subspace of scalar-tensor theories emerging from massive gravity, the only stable field configuration created by static spherically symmetric sources was one with cosmological asymptotics. Moreover, these backgrounds  were shown to be sub-luminal everywhere in the space; in contrast to the common believe that these theories are necessarily superluminal in the vicinity of a static source. In this work we complete that analysis by extending it to cover the whole parameter space of these scalar-tensor theories. We find that the stability argument renders the asymptotically flat backgrounds unrealizable, forcing once again for cosmological asymptotics. In the case of pressureless sources these backgrounds are stable. However, they get destabilized in the presence of positive pressure, larger than a critical density. Even on the self-accelerated background, on which the scalar mode decouples from sources, in the region occupied by the source it acquires an elliptic equation of motion. Therefore, we conclude that the only parameter space which is not ruled out, by solar system measurements, is the one considered in Berezhiani {\it et al.} (arXiv:1302.0549), namely the one for which the scalar and tensor modes can be diagonalized via local transformations.

%Furthermore, in order to study the strong coupling issue,
We also reinvestigate the scale at which perturbation theory breaks down in a general Galileon theory. We show that the Vainshtein mechanism successfully redresses the strong coupling scale to a small one, just like in the cubic Galileon, despite the cancellations occurring in the special spherically symmetric case. We emphasize that even if these tests were performed at scales at which perturbation theory broke down, these could not be interpreted as a lower bound for the graviton mass.

\end{abstract}

\newpage

\section{Introduction and Summary}

There exists a two-parameter family of theories of massive gravity that propagate five degrees of freedom in four dimensions \cite{dRG,dRGT}. In the decoupling limit, it gives rise to a fascinating class of scalar-tensor theories. There is a one-parameter sub-class of these theories, for which the scalar mode can be completely decoupled from the tensor  using the invertible field redefinition $h\mn\to \bar h\mn+\pi \eta\mn +\alpha \p_\mu \pi \p_\nu \pi/\Lambda_3^3$, \cite{dRG}. As a result of this de-mixing, the longitudinal mode of the graviton becomes described by the so-called quartic `Galileon' model \cite{Rattazzi2}\footnote{This type of theories were first discovered in \cite{Rattazzi1} in the context of the DGP scenario \cite{DGP}, see also \cite{deRham:2010gu}}, supplemented with a novel coupling to the matter stress-tensor $\partial_\mu \pi \partial_\nu \pi T^{\mu\nu}$, where $\pi$ denotes the helicity-$0$ mode of the graviton and $T\mn$ denotes the matter energy-momentum tensor, \cite{dRG}.
After partial diagonalization, the Lagrangian describing the scalar sector reduces to
\beq
\mathcal{L}_\pi=\frac{3}{2}\pi\Box\pi+\frac{3}{2} \frac{\alpha}{\Lambda_3^3}
(\partial\pi)^2\Box \pi+\frac{1}{2}\frac{\alpha^2}{\Lambda_3^6}
(\partial\pi)^2\left((\partial\partial \pi)^2-(\Box\pi) ^2\right)\nonumber \\
+\frac{1}{\mpl}\pi T+\frac{\alpha}{\mpl\Lambda_3^3}
\partial_\mu \pi \partial_\nu \pi T^{\mu\nu},
\label{diag}
\eeq
where  $\Lambda_3\equiv (\mpl m^2)^{1/3}$ is the strong coupling scale and $\alpha$ is the free real parameter of this sub-class of theories.

The thorough study of the spherically symmetric configurations \cite{BCG} (for related work on spherical symmetric solutions see \cite{Volkov:2013roa}) showed that the stability of the spherically symmetric configurations in the presence of the static source forces the free parameter $\alpha$ to be positive. Otherwise, the last term of \eqref{diag} gives rise to a ghost-like kinetic term in the high density regions. Furthermore, it was shown in \cite{BCG} that, for $\alpha>0$, the theory does not admit asymptotically flat classical solutions. Instead, the screened $\pi$ configuration at short distances (within the Vainshtein region \cite{Arkady}, see also \cite{Babichev:2013usa} for a recent review on the Vainhstein mechanism) matches a cosmological background at large distances. We emphasize that this matching effect is not related to the last term of \eqref{diag} and would be present even in its absence. In addition, it was explicitly shown that these backgrounds are subluminal, as opposed to the common believe that in massive gravity the configurations recovering GR at short distances necessarily exhibit superluminal propagation\footnote{We emphasize however that the `issue' of superluminality in massive gravity has not (yet) been connected to that of acausality in any rigorous way. Configurations on which closed time-like curves could form seem to live beyond the regime of validity of the effective theory,  \cite{Burrage:2011cr}.}.

In the present work we would like to complete the analysis of \cite{BCG} by extending it to the full parameter space of the decoupling limit of massive gravity. In general, there is a  nonlinear mixing term between the helicity-0 and -2 modes which cannot be removed by a local field redefinition. This makes the assay slightly more involved and gives rise to a qualitative change in the conclusions.
Namely, for generic parameters of the theory, we show that asymptotically flat backgrounds created by static spherically symmetric sources exhibit a gradient instability. This should be contrasted with the case considered in \cite{BCG}; where asymptotically flat configurations were unstable as well, although the instability was ghost-like.

The presence of this gradient instability in the general case forces us to give up asymptotic flatness, and we instead focus on the  self-accelerated background configuration. The latter was already shown to be interesting for its unique property to decouple the $\pi$ mode from matter (at leading order), \cite{deRham:2010tw,Wyman}, not to mention the obvious phenomenological significance of such backgrounds. The absence of a direct conformal coupling $\pi T$ on top of this self-accelerating solution implies that no Vainshtein mechanism or other screening effect ought to be invoked. Furthermore, after using Einstein's equations, this decoupling leads to the following contributions to the kinetic term of the scalar mode in the vicinity of the source
\beq
\L_{\pi} \supset -\(\eta^{\mu\nu}-\frac{1}{\mpl \Lambda_3^3}T^{\mu\nu}\)\partial_\mu\pi \partial_\nu\pi\,.
\label{ds}
\eeq
The examination of  \eqref{ds} shows that, in case of a dust-like source, both contributions to the kinetic term $\dot{\pi}^2$ are healthy; hence, the scalar mode is not a ghost. However, it is straightforward to see that in case of the positive pressure the second term of \eqref{ds} gives rise to a negative contribution to the gradient energy. Moreover, if the pressure is larger than the critical density $\rho_c=\mpl \Lambda_3^3$, something which is common for astrophysical sources or indeed even for the atmosphere! this negative contribution overcomes the positive one, making the background unstable. This argument can be extended to other asymptotic cosmologies, as long as we recover General Relativity at short distances the same instability will arise.

Therefore, if the scalar-tensor theory considered here were to 
reproduce all the important effects of the full theory,  then we would conclude 
that the only phenomenologically viable theory  of massive gravity is the 
one with $\beta=0$ and $\alpha>0$.  However, the scalar-tensor theory may not be necessarily 
capturing all the important properties of massive gravity. In particular, 
the presence of a nonzero  time-like component of the helicity-1 field ($A_0\neq 0$) 
may give rise to a new class of spherically symmetric solutions not captured 
by the scalar-tensor sector.

In the second part of this manuscript, we investigate the strong coupling issue in theories such as massive gravity and Galileons.
The fact that the Vainshtein mechanism relies on irrelevant operators (in the Wilsonian sense) to be important raises the question of the control one has over perturbation theory. As previously investigated, \cite{Rattazzi1} over a background configuration for the scalar field
$\pi=\pi_0+\delta \pi$, the strong coupling scale gets redressed, symbolically  $\Lambda_{\rm redressed}\sim \Lambda_3 (\p^2 \pi_0/\Lambda_3^3)^a$, with a power $a$ depending on the exact model. In the cubic Galileon that arises in DGP, $a=1/2$, while in models with higher order Galileon interactions, one could have $a =2/3$. This redressing allows to raise the strong coupling scale to $\sim( 1{\rm cm})^{-1}$ in the cubic Galileon case, however this scale is still low enough to wonder what happens at energy scales above that. Furthermore in the quartic Galileon and some models of massive gravity, it has been argued that the redressing only raises the strong coupling scale to $\sim (0.4{\rm km})^{-1}$ making the theory fully non-perturbative at extremely low energy scales, \cite{1211.6001}. We show here that the difference between the cubic Galileon and the quartic Galileon/massive gravity is actually not as pronounced as previously found, and  the redressed strong coupling scale in the quartic Galileon/massive gravity is actually closer to $\sim (30{\rm cm})^{-1}$, \ie at most one order of magnitude below the cubic Galileon.

Putting this subtlety aside, one can raise the interesting question of what happens when reaching energy scales higher than the redressed strong coupling scale, and perturbation theory runs out of control. Following Vainshtein's original argument (rather than its specific realization within a Galileon theory), we do expect the effect of the graviton mass and in particular the effect of the helicity-0 mode to become smaller and smaller as one gets to higher and higher energies.
%In this manuscript we discuss how the phenomena is `expected', even though an explicit derivation is beyond the scope of this work.
We emphasize however that there is no sense in which one could use the break of perturbation theory as a lower bound for the graviton mass as has been done in the literature \cite{1211.6001}.

The paper is organized as follows. Section 2 is dedicated to the description of the framework. In section 3 we analyze the spherically symmetric configuration and  study its stability both for asymptotically flat backgrounds and for the self-accelerated one. We also discuss the stability for more general cosmologies. Finally, we reexamine the strong coupling issue in section 4, and open the discussion for a deeper understanding in section 5.

As we were finalizing this paper, it came to our attention that the related work was being conducted in \cite{K}; which has some overlap with our work.

\section{The Theory}

The scalar-tensor sector of massive gravity, in the decoupling limit, is described by the following Lagrangian density
\beq
\mathcal{L}= -\frac{1}{2} h^{\mu\nu}\mathcal{E}_{\mu\nu}^{\alpha\beta} h_{\alpha\beta} + h^{\mu\nu}X^{(1)}_{\mu\nu}+\frac{\alpha}{\Lambda^{3}_{3}}h^{\mu\nu}X^{(2)}_{\mu\nu} +\frac{\beta}{\Lambda^{6}_{3}}h^{\mu\nu}X^{(3)}_{\mu\nu}+\frac{1}{ \mpl}h^{\mu\nu}T_{\mu\nu},
\label{lagr}
\eeq
where, we have denoted the helicity-$\pm2$,$0$ modes by $h\mn$ and $\pi$ respectively.

The three identically conserved symmetric tensors  $X^{(n)}_{\mu\nu}[\Pi]$ depend on second derivatives of the helicity-0 field $\Pi_{\mu\nu}\equiv \partial_\mu \partial_\nu \pi$ in the following way,
\begin{align}
X^{(1)}_{\mu\nu}=-\frac{1}{2}{\varepsilon_{\mu}}^{\alpha\rho\sigma}{{\varepsilon_\nu}^{\beta}}_{\rho\sigma}\Pi_{\alpha\beta}, \quad  \nonumber \\
X^{(2)}_{\mu\nu}=\frac{1}{2}{\varepsilon_{\mu}}^{\alpha\rho\gamma}{{\varepsilon_\nu}^{\beta\sigma}}_{\gamma}\Pi_{\alpha\beta}\Pi_{\rho\sigma}, \nonumber \\
X^{(3)}_{\mu\nu}={\varepsilon_{\mu}}^{\alpha\rho\gamma}{{\varepsilon_\nu}^{\beta\sigma\delta}}\Pi_{\alpha\beta}\Pi_{\rho\sigma}\Pi_{\gamma\delta}.
\nonumber
\end{align}
The Lagrangian \eqref{lagr} is invariant under the linear diffeomorphisms $\delta h_{\mu\nu} = \partial_{\mu}\zeta_{\nu} +
\partial_{\nu}\zeta_{\mu}$ up to a total derivative, while being {\bf exactly} invariant under the global Galilean symmetry $\delta \pi = v_\mu x^\mu$.

Recently, it has been shown \cite{dRGHP2} that the decoupling limit of massive gravity exhibits the same non-renormalization properties as the Galileon theory \cite{Rattazzi2,Rattazzi1}. Namely, the coefficients $\alpha$ and $\beta$ do not get radiatively corrected, within the effective theory.

In \cite{BCG}, it was found that the $\beta=0$ parameter subspace of \eqref{lagr} does not possess a stable asymptotically flat solutions in the presence of the spherically symmetric static sources. Instead, the stable screened $\pi$ configuration inside the Vainshtein region is smoothly matched to the cosmological backgrounds (with various equations of state) at large distances. In this work, we would like to perform the similar analysis for the rest of the parameter space (\ie $\beta\neq 0$), which is qualitatively different from its $\beta=0$ counterpart. In particular, in the present case the mixing between the different helicity states cannot be undone by means of a local field redefinition. This structure is so far characteristic to massive gravity \cite{dRGT} and has not been observed in other modifications of General Relativity.

\section{Static Spherically-Symmetric Configurations}

Throughout this work we consider a star-like source of finite size $R$ and uniform density $\rho$. Without loss of generality, the static spherically symmetric configuration can be found by assuming the following ansatz for the metric perturbations around a Minkowski space-time in spherical coordinates
\beq
h_{00}=a(r); \qquad h_{ij}=f(r)\delta_{ij}.
\label{anz}
\eeq
The most general ansatz (up to Galilean transformations) for the helicity-$0$ mode, which leads to the static spherically symmetric metric \eqref{anz} (see appendix A), is given by
\beq
\pi=\frac{c}{2}\Lambda_3^3t^2+\pi_0(r)\,.
\label{pianz}
\eeq
After integrating Einstein's equations, \ie the equations for $h_{\mu \nu}$, once and using vanishing initial conditions in the origin we obtain
\beq
\label{1}
&&rf^\prime=-\frac{2M}{\mpl r}+\Lambda_3^3 r^2\lambda(1-\alpha\lambda-2\beta \lambda^2) ,
\\&& ra^\prime=-\frac{2M}{\mpl r}+\Lambda_3^3 r^2(c-\lambda(1+2\alpha c)-6\beta c \lambda^2-2\beta\lambda^3)\,.
\label{2}
\eeq
Following the same procedure for the longitudinal mode we arrive at the following equation
\beq
&&3(1+2\alpha c)\lambda-6(\alpha+\alpha^2 c -4\beta c)\lambda^2+2 (\alpha^2-4\beta-20\alpha\beta c)\lambda^3 - 60\beta^2 c \lambda^4 - 12\beta^2\lambda^5 \nonumber
\\&&= \left\{ \begin{array}{ll}
2\left(\frac{r_*}{r}\right)^3 (1+2\alpha c+12\beta c \lambda+6\beta\lambda^2)+c
& \mbox{Outside the source} \\ 2\left(\frac{r_*}{R}\right)^3 (1+2\alpha c+12\beta c \lambda+6\beta\lambda^2)+c
& \mbox{Inside the source} \\ \end{array}
\right .\,,
\label{3}
\eeq
where we have defined
\beq
\lambda\equiv \frac{\pi_0'}{\Lambda^3_3 r}, \qquad {\rm and } \qquad r_*\equiv \( \frac{M}{\mpl \Lambda_3^3} \)^{1/3}\,,
\eeq
$r_\star$ being the Vainshtein radius and $\pi'_0\equiv\p_r\pi_0(r)$.
The classical backgrounds in the time independent case ($c=0$) have been previously studied in \cite{GigaDato,Koyama3}. It has been established that for generic parameters \eqref{3} possesses two types of solutions inside the Vainshtein region. In particular, since the factor $(r_*/r)^3$ is large at short distances, \eqref{3} requires either (i) $\lambda\gg 1$ or (ii) the vanishing coefficient of $(r_*/r)^3$. Moreover, it has been shown that the case (i) corresponds to a screened gravitational field at short distances, hence contradicting empirical data.

In this work we study  the general class of spherically symmetric configurations \eqref{pianz}, created by static sources, against instabilities.

\subsection{Asymptotic Flatness}

As it has been already mentioned for generic $\alpha$ and $\beta$ the equation of motion for the longitudinal mode has two types of solutions at short distances, $r\ll r_*$. One of these solutions is
\beq
\lambda=-\beta^{-1/3}\frac{r_*}{r},
\eeq
which corresponds to a screened gravitational field, \ie a field for which the Newtonian $1/r$ behavior is absent, at short distances even for nontrivial $c$; as it is easy to see from \eqref{1} and \eqref{2}. We dismiss this solution for obvious reasons\footnote{Moreover, for $c=0$ this solution matches the cosmological background, as it was noted in \cite{GigaDato}.}.

The other solution corresponds to the case when the coefficient of $(r_*/r)^3$ in \eqref{3} vanishes on the background, that is
\beq
1+2\alpha c+12\beta c \lambda+6\beta\lambda^2=0.
\label{GR}
\eeq
Which evidently gives us a solution, with $\lambda=const$; hence, recovering GR at short distances with high precision (see \eqref{1} and \eqref{2}; the normal $1/r$ behavior of gravity at large distances precisely cancels). This serves as a motivation for studying the stability of the generic background with constant $\lambda$. Within the Vainshtein region the leading contribution to the gradient energy, to the second order in fluctuations reads as follows
\beq
\mathcal{L}^{(2)}=- 3\beta (c+\lambda)\left(\frac{r_*}{r}\right)^3\left[ 2 (\partial_r \delta\pi)^2-(\partial_\Omega \delta\pi)^2 \right].
\label{tachyon}
\eeq
Notice the relative minus sign between the radial and angular terms\footnote{Here, we would like to emphasize that \eqref{tachyon} comes  from the $h^{\mu\nu}X^{(3)}\mn$ term of \eqref{lagr}; the other contributions to the quadratic gradient term are subdominant.}. As a result, we deduce that the only way to avoid the gradient instability is to have $\lambda\simeq-c$ at short distances. This condition, combined with \eqref{GR} leads to
\beq
c=\frac{\alpha\pm \sqrt{\alpha^2+6\beta}}{6\beta}.
\eeq
If we further require asymptotic flatness, we are led to the following system of conditions
\beq
\lambda_\infty(1-\alpha\lambda_\infty-2\beta \lambda_\infty^2)=0,\\
c-\lambda_\infty(1+2\alpha c)-6\beta c \lambda_\infty^2-2\beta\lambda_\infty^3=0,
\eeq
where $\lambda_\infty$ denotes the value of $\lambda$ at spatial infinity. It is easy to show that these conditions possess a nontrivial solution only in the parameter space
\beq
\beta=-\frac{\alpha^2}{8}.
\eeq
However, according to \cite{pain}, in this parameter subspace the flat vacuum with $\lambda=-c$ is infinitely strongly coupled.

Hence, the analysis of this section leads us to the conclusion that the theory under consideration does not possess a stable asymptotically flat solution sourced by a static and spherically symmetric source.

\subsection{Asymptotically de Sitter}

Having failed to introduce a source on flat space, it is natural to ask what happens when we substitute the flat space by a self-induced de Sitter vacuum.
We look for the de Sitter space solution in static slicing for which the ansatz adopted in the previous section is applicable here as well.

Therefore, we are looking for the spherically symmetric solutions to \eqref{3} with de Sitter asymptotic. This means that at large distances the effective energy momentum tensor, coming from the mass term of the graviton, will have the equation of state of the cosmological constant. As a result, using \eqref{1}-\eqref{3}, we arrive at
\beq
c=-\lambda(r\rightarrow \infty)=-\frac{-\alpha\pm\sqrt{\alpha^2+6\beta}}{6\beta}.
\eeq
Then, the eq.\eqref{3} at arbitrary distances factorizes and takes the following form
\beq
(\lambda+c)P(\lambda)=12 \beta \left(\frac{r_*}{r}\right)^3 (\lambda+c)^2,
\eeq
where, $P(\lambda)$ is the polynomial of the fourth order; which does not vanish when $\lambda=-c$, unless $\beta=0$, $\beta=-\alpha^2/6$ or $\beta=-\alpha^2/8$. The first possibility leads to a ghost-like instability of the de Sitter space \cite{deRham:2010tw}, in the second case the helicity-$0$ loses its kinetic term, while the third one necessarily leads to an asymptotically flat background (which has already been discarded due to infinite strong coupling).  This implies that the only solution for the generic parameters (except the above-mentioned ones) is the one with the trivial (source independent) $\pi$ profile around the source
\beq
\pi=-\frac{c}{2}\Lambda_3^3  \, x_\mu x^\mu.
\eeq
This may be traced to the fact that, according to \cite{deRham:2010tw}, on de Sitter space the helicity-$0$ mode does not have the kinetic mixing with the helicity-$\pm 2$. Namely, the Lagrangian density re-calculated on the self-accelerated background is given by
\beq
\mathcal{L}=-\frac{1}{2} h^{\mu\nu}\mathcal{E}_{\mu\nu}^{\alpha\beta} h_{\alpha\beta}+\kappa(\alpha,\beta)\frac{H^2\mpl}{\Lambda^3}\pi\Box\pi+\frac{\kappa(\alpha,\beta)}{\Lambda^3}h^{\mu\nu}X^{(2)}\mn  \nonumber\\
-3\beta\frac{H^2\mpl}{\Lambda^6}(\partial\pi)^2\Box \pi+\frac{\beta}{\Lambda^6}h^{\mu\nu}X^{(3)}\mn+\frac{1}{\mpl}h^{\mu\nu}T_{\mu\nu}.
\label{deSitter}
\eeq
Here, $h\mn$ and $\pi$ denote the deviations from the de Sitter background for the tensor and scalar modes respectively. The curvature of the background is given by $H^2\propto m^2$ and $\kappa$ is some function of the parameters. The exact expressions for them are inessential for current discussion and in case of interest the reader is directed to \cite{deRham:2010tw}.

Again, the only asymptotically decaying solution to the equations of motion, following from \eqref{deSitter}, is $\pi=0$. While $h\mn$ satisfies the linearized Einstein's field equations
\beq
G\mn^{(1)}=\frac{1}{\mpl}T\mn.
\label{einst}
\eeq
The problem in case of certain sources arises from the third term of \eqref{deSitter}, which by integration by parts can be rewritten in the following form
\beq
\frac{\kappa(\alpha,\beta)}{\Lambda^3}G\mn^{(1)}\partial^\mu \pi \partial^\nu \pi.
\label{dom}
\eeq
In case of the sources smaller than their Vainshtein radius this term will overwhelm the kinetic term inside the sources, precisely like in \cite{BCG}. Hence the $\pi$ becomes a ghost unless $\kappa>0$; surprisingly, this is exactly what we need in order to have the healthy kinetic term outside of the source as well \eqref{deSitter}.

Therefore, we conclude that the theory admits the presence of the arbitrarily dense objects on de Sitter space. However, this is so only for pressure-less sources; otherwise an extra caution is in order. Namely, if the source has a pressure, larger than the critical density, then the dominant contribution to the quadratic term comes entirely from \eqref{dom}. Moreover, in case of the positive pressure this contribution has a wrong sign (as it follows from \eqref{einst}) in front of the gradient term, leading to the gradient instability of the configuration. Unfortunately, most of the localized sources have sufficiently large pressure to realize this instability; even the atmosphere around the earth has the pressure $10^{14}$ times larger than the critical density.

\subsection{Other Cosmologies}

We would like to start this section by pointing out that, up to this point the situation is qualitatively similar to the $\beta=0$ case, see \cite{BCG}. Namely, the presence of natural sources seems to be destabilizing the asymptotically flat, as well as asymptotically self-accelerated spaces (in case of $\beta=0$ there was no stable self-accelerated space to begin with, see \cite{deRham:2010tw}). However, in that case the situation has changed once we allowed for cosmological backgrounds with nontrivial equation of state \cite{BCG}. In this section we would like to explore this possibility for the $\beta\neq 0$ parameter space.

First, we reiterate that the stability of the background forces upon us the condition $\lambda\simeq -c$ within the Vainshtein region (see eq.\eqref{tachyon}). As a result, the leading contribution to the quadratic (in perturbations) Lagrangian vanishes. In order to find the next order, non-vanishing result, we have to find the corrections to the background itself. The correction to $\lambda$ is parameterized as $\lambda=-c+\delta \lambda$; where, as it is easy to see from \eqref{3}, $\delta \lambda$ is of order of $(r/r_*)^3$. While the expression for the metric degrees of freedom at next to leading order is obtained by substituting $\lambda=-c$ to the right hand side of \eqref{1} and \eqref{2}; leading to the correction $\sim \Lambda_3^3 r^2$. In other words, the background next to the leading order has the following form
\beq
\lambda&\simeq& -c+\text{O}\left[\left(\frac{r}{r_*}\right)^3\right], \\
\label{fcor}
f&\simeq& \frac{M}{\mpl r}+\frac{\Lambda_3^3r^2}{2}~c~(-1-\alpha c+2 \beta c^2),\\
\label{acor}
a&\simeq& \frac{M}{\mpl r}-\Lambda_3^3 r^2 ~c~(-1-\alpha c+2 \beta c^2).
\eeq
By simple analysis of these corrections we conclude that the leading contribution to kinetic terms of fluctuations comes from the background
\beq
\lambda=-c,
\label{des}
\eeq
while for $f$ and $a$ we should take into account the corrections present in \eqref{fcor} and \eqref{acor}. To this end, the background relevant for the quadratic terms is similar to the one for de Sitter space (see the previous subsection). With the distinction that in the case of de Sitter space the background \eqref{des} was exact, while in the current case it is merely the leading expression for the background. Despite this minor distinction, since \eqref{des} is the relevant piece of the background at short distances, the conclusions of the previous section apply here as well. Namely, the background will exhibit the instability either inside the source or right outside of it. It should also be mentioned that the mixing between the scalar and tensor modes of the graviton are irrelevant, since to the leading order we will have the pure $\pi$ kinetic term, while the $h\pi$ mixing appears only in the sub-leading approximation.

We would like to conclude by noting that in case of $\beta\ll \alpha$ there seems to be another branch of solutions at short distances. Namely, one could think of  having $\beta$ so small that the right-hand side of \eqref{3} could have been neglected at short distances; such a small values for the parameter are technically natural since it does not get renormalized by quantum loops \cite{dRGHP2}, so this would not represent a fine-tuning of the parameters. For simplicity, let us concentrate on $c=0$ and $\alpha= 1$ as it is quite straightforward to generalize the argument. This branch of solutions requires $\lambda\gg 1$, which under the assumption that the left hand side is negligible gives
\beq
2\lambda^3\simeq 12\beta^2\lambda^5 \qquad \Rightarrow \qquad \lambda\simeq \pm\frac{1}{\sqrt{6}\beta}.
\eeq
However, in order this assumption about the right hand side to be valid, we need to make sure that
\beq
\frac{1}{\beta^2}\gg \left(\frac{r_*}{R}\right)^3.
\eeq
Taking into account \eqref{1}, \eqref{2} and the expressions given above, it is easy to see that this branch of solutions gives an unacceptably large deviation from GR. Hence, it is ruled out on phenomenological grounds.

\section{Strong Coupling for general Galileons}

In the previous section, we have shown that for massive gravity, stability of the spherically symmetric solutions imposes a non-flat, \ie cosmological, asymptotic behaviour. This is however not necessarily the case for a more general Galileon theory where the coefficients between the different operators are relaxed and there is no mixing with the helicity-2 mode of the graviton. In this section we investigate the redressing of the strong coupling scale on spherically symmetric configurations and discuss the small departure from spherical symmetry. We start with a generic Galileon setup  and compute the redressing of the strong coupling scale both in the cubic and the quartic Galileon. Within the Vainshtein regime, the stable solution in massive gravity covered by $\alpha>0$ is essentially a special case of the quartic Galileon (although the asymptotics outside the Vainshtein region are different).

\subsection{Asymptotically Flat Spherically Symmetric Galileons}

To start with, we consider a quartic Galileon Theory in four dimensions
\ba
S=\int \d^4 x \(-\frac 32 \sum_{i=2}^4 \frac{c_i}{\Lambda_3^{3(i-2)}}\mathcal{L}_i\)+ \frac{\pi}{\mpl}T\,,
\ea
where $T$ is the trace of the stress-energy tensor of any external sources and the Galileon Lagrangians are given by
\ba
\mathcal{L}_2&=&(\p \pi)^2\\
\mathcal{L}_3&=&(\p \pi)^2 \Box \pi\\
\mathcal{L}_4&=&(\p \pi)^2 \((\Box \pi)^2-(\p_\mu\p_\nu \pi)^2\)\,.
\ea
In massive gravity (with $\beta=0$), $c_2=1$, $c_3=-\alpha$ and $c_4=\alpha^2/3$, but we leave the $c_3$ and $c_4$ coefficients arbitrary for now and simply set $c_2=1$. We are mainly interested in the Vainshtein screening of this Galileon due to the Earth, so as first approximation, the background source may be considered as spherically symmetric, $T=-M_\Earth \delta^{(3)}(r)+\delta T$, leading to a background configuration for the Galileon field which is also spherically symmetric
\ba
\pi(t,\vec r)= \pi_0(r)+\frac{1}{\sqrt{3}}\phi (t, \vec r)\,,
\ea
where the background field satisfies the simple algebraic equation,
\ba
\frac{\pi_0'(r)}{r}+\frac{2c_3}{\Lambda_3^3}\(\frac{\pi_0'(r)}{r}\)^2
+ \frac{2 c_4}{\Lambda_3^6}\(\frac{\pi_0'(r)}{r}\)^3=\frac{1}{12 \pi}\frac{M_\Earth}{\mpl}\frac 1{r^3}\,.
\ea
If the quartic Galileon is present, $c_4\ne 0$, then close to the source, (at small $r$), the last term on the left hand side dominates and the background solution takes the form
\ba
\pi_0'(r)=\La^2\Big[\(\frac{1}{12 \pi c_4}\frac{M_\Earth}{\mpl}\)^{1/3}+\mathcal{O}(\La r)\Big]\,,
\ea
while if $c_4=0$, the background field acquires a different profile,
\ba
\pi_0'(r)=\La^2\Big[\(\frac{1}{12 \pi c_3}\frac{M_\Earth}{\mpl}\frac{1}{\La r}\)^{1/2}+\mathcal{O}(\La r)\Big]\,.
\ea
The perturbations around this background configuration evolve with the following kinetic matrix $Z^{\mu\nu}$,
\ba
\mathcal{L}_{\phi}=-\frac 12 Z^{\mu\nu}(r)\p_\mu \phi \p_\nu \phi+ \cdots\,,
\ea
with
\ba
\label{eq:def-Zrr}
&& Z_{rr}(r)  = 1+\frac{4c_3}{\La^{3}}\frac{\pi_0'}{r}+\frac{6c_4}{\La^{6}}\frac{\pi_0'^{2}}{r^{2}}\\
\label{eq:def-Ztt}
&& Z_{tt}(r)  =  \frac{1}{3r^{2}}\frac{\d }{\d  r} \left[ r^{3}\left(1+\frac{6c_3}{\La^{3}}\frac{\pi_0'}{r}+\frac{18c_4}{\La^{6}}\frac{\pi_0'^{2}}{r^{2}}\right)\right]\\
\label{eq:def-ZOO}
&& Z_{\Omega\Omega}(r)  =  \frac{1}{2r}\frac{\d }{\d  r}\left[r^{2}\left(1+\frac{4c_3}{\La^{3}}\frac{\pi_0'}{r}+\frac{6c_4}{\La^{6}}\frac{\pi_0'^{2}}{r^{2}}\right)\right]\,.
\ea
When the quartic Galileon is present, one can see immediately that the leading contribution to the angular direction vanishes,
$Z_{tt}\sim Z_{rr}\sim \(\La r\)^{-2}\(M_\Earth/\mpl\)^{2/3}$ while $Z_{\Omega \Omega}\sim \(\La r\)^{-1} \(M_\Earth/\mpl\)^{1/2}$. This leads to a few subtleties in the quartic Galileon but as we will see later, the resulting redressed strong coupling scale is nevertheless barely affected by this subtlety. The reason for that is that the same cancelation responsible for the hierarchy $Z_{\Omega \Omega}\ll Z_{tt}$ is also responsible for cancelling the leading contribution to the operator that would naively arise at the lowest energy scale. As a result the redressed strong coupling scale is larger than naively anticipated.

\subsection{Cubic Galileon}

\subsubsection{Redressing from the Earth}

We start by focusing on the cubic Galileon and set $c_3=1/3$ for simplicity. In that case
\ba
Z_{tt}\sim Z_{rr}\sim Z_{\Omega \Omega}\sim \(\frac{M_\Earth}{4 \pi\mpl}\)^{1/2}\frac{1}{\(\La r\)^{3/2}}\equiv Z_\Earth\,.
\ea
The canonically normalized field is then
\ba
\hat \phi \sim \sqrt{Z_\Earth}\phi\,,
\ea
leading to the cubic interaction
\ba
\mathcal{L}^{(3)}_\phi=\frac{1}{\La^3}\(\p \phi\)^2 \Box \phi
=\frac{1}{\La^3 Z_\Earth^{3/2}}\(\p \hat \phi\)^2 \Box \hat\phi
=\frac{1}{\LE^3}\(\p \phi\)^2 \Box \phi\,.
\ea
So the redressed coupling scale due to the screening of the Earth is given by
\ba
\LE=\La \sqrt{Z_\Earth}\sim \La \(\frac{M_\Earth}{4 \pi\mpl}\)^{1/4}\frac{1}{\(\La r\)^{3/4}}\,.
\ea
Taking the strong coupling scale to be that associated with an infrared modified theory such as DGP or massive gravity for which $\La = (H_0^2 \mpl)^{1/3}$ where $H_0\sim 10^{-33}$eV, we have $\La\sim(1120{\rm km})^{-1}$, and so at the surface of the Earth the redressed scale is (as previously found in \cite{RattazziNicolis})
\ba
\LE\sim  10^7 \La \sim \(4\, {\rm cm}\)^{-1}\,.
\ea

\subsubsection{Redressing from Nearby Sources}

When testing the Newton's law  using torsion balance at submillimeter scales, the experiment itself and nearby sources can further contribute to the screening of the Galileon field.
%For definiteness let us consider that the pendulum itself weighs $M_e\sim10$kg and is $\rho_e\sim5$cm long. (some of the latest generation pendulum have been using $4\times 8.14$kg masses within separated by $\sim20$cm. The recent Eot-Wash experiments use instead two disks, 5cm diameter, 2mm thick, a few $\mu$m apart, about $100$g each, but i dont know how heavy the vacuum chamber and other apparatus are).
%
For concreteness, let us consider a source of mass $M_e$ (which could symbolize the experiment itself or a nearby source) localised a distance $\rho_e$ from the core of the experiment.
The coupling to that new source leads to a new field configuration $\pi_e(\rho)$ on the top of that $\pi_0(r)$, given by
\ba
\pi_e'(\rho)=\LE^2\(\frac{1}{4 \pi}\frac{M_e}{\sqrt{Z_\Earth} \mpl}\frac{1}{\LE \rho}\)^{1/2}\,,
\ea
where $\rho$ is the distance from the source $M_e$.
Going through the same analysis presented previously, the redressing of the strong coupling scale due to the mass $M_e$ is then
\ba
\Lambda_e=\LE  \(\frac{M_e}{4 \pi\sqrt{Z_\Earth}\mpl}\)^{1/4}\frac{1}{\(\LE \rho_e\)^{3/4}}\,.
\ea
As a possible example, if we consider that the local effects could be mimicked by a mass of $10$kg localized $1$cm away from the center of the experiment then,
\ba
\Lambda_e\simeq 4 \ \Lambda_\Earth\,.
\ea
In itself this is not a huge enhancement, but it simply serves to show that as one goes within the apparatus itself, its different components help raising the scale, and screening the force.

\subsection{Quartic Galileon - Massive Gravity}

In massive gravity, when $\beta=0$, the decoupling limit resembles that of a quartic Galileon with an additional coupling to matter. As we will discuss in section \ref{sec:BiggerPicture}, in that case, the only relevant scale in the decoupling limit is not $\Lambda_3=(m^2\mpl)^{1/3}$ but rather
\ba
\Lambda\equiv\frac{\Lambda_3}{\alpha^{1/3}}\,.
\ea
Since the coefficient $\alpha$ does not renormalize \cite{dRGHP2}, $\alpha$ can in principle depart significantly from unity, hence disentangling the strong coupling scale $\Lambda$ which appears in the decoupling limit, with the graviton mass $m$. In what follows we will thus set $c_4=\alpha^2/3$ as is the case in massive gravity and work with the scale $\Lambda$ which can in principle be independent from the graviton mass.

\subsubsection{Angular Subtleties}

As mentioned previously, in the {\it purely} spherically symmetric case, when in the vacuum $Z_{\Omega \Omega}\ll Z_{rr}\sim Z_{tt}$, which leads to a few subtleties in treating this case. In what follows we consider the case where
\ba
&&Z_{rr}\sim Z_{tt} \equiv Z_\Earth =\frac{6}{\Lambda^2 R_\Earth^2}\(\frac{1}{12\sqrt{3}\pi}\frac{M_\Earth}{\mpl}\)^{2/3} \\
{\rm and}\hspace{10pt}
&& Z_{\Omega \Omega}= \e\  Z_\Earth\,,
\ea
with $\e \ll 1$.

$\bullet$ If we only consider the effect from the Earth itself, and are interested in the redressed scale in the vacuum outside the Earth, then as seen earlier,
\ba
\label{naive epsilon}
\e\sim \frac 19  \frac{\Lambda R_\Earth}{\(\frac{1}{12\sqrt{3} \pi}\frac{M_\Earth}{\mpl}\)^{1/3}}\sim  {\rm few}\times 10^{-11}\(\frac{\Lambda}{(H_0^2\mpl)^{1/3}} \)\,.
\ea
$\bullet$ However, the Earth itself is not perfectly spherically symmetric, and just taking into account the flatness of the Earth, (which is of the order of $\delta\sim 0.0033$), we would have instead more realistically $\e\sim \delta^2\sim 10^{-5}$. Furthermore, the presence of other sources near the experiment itself will completely break the symmetry and more realistically, we would expect $\e\sim 1$ at the level of the experiment itself. However for consistency we keep $\e$ as an arbitrary parameter for now, with $\e \ll 1$.

Since the gradient along the angular direction are not redressed with the same scale as along the radial direction, we first need to rescale the angular directions as, \cite{1211.6001}
\ba
(t,r, \theta, \varphi)=(\hat t, \hat r, \e^{1/2}\hat \theta,\e^{1/2} \hat \varphi)\,,
\ea
this is not something one could do globally, but if we are only interested in what happens in a small region of space near say an experiment, and not for all angles, the rescaling can be done locally. The kinetic term is then of the form
%\ba
%\int \d^4x {\mathcal L}_{\phi}&=&\int \d^4 x \(-\frac 12 Z^{\mu\nu}(r)\p_\mu \phi \p_\nu \phi+ \cdots\)\nn\\
%&\sim&\int \d^4 \hat x\ \e \(-\frac 12 Z_\Earth \hat \p_\mu \phi \hat \p_\nu \phi+ \cdots\)\nn\\
%&\sim&\int \d^4 \hat x \(-\frac 12 \hat \p_\mu \hat \phi \hat \p_\nu \hat \phi+ \cdots\)\,,
%\ea
\ba
\int \d^4x {\mathcal L}_{\phi}=\int \d^4 x \(-\frac 12 Z^{\mu\nu}(r)\p_\mu \phi \p_\nu \phi+ \cdots\)
\sim \int \d^4 \hat x \(-\frac 12 \hat \p_\mu \hat \phi \hat \p_\nu \hat \phi+ \cdots\)\,,
\ea
with $\hat \phi=\sqrt{\e Z_\Earth}\, \phi$. In terms of the canonically normalized field $\hat \phi$, the quartic and cubic Galileon leads to cubic interactions of the form (focusing only on the interactions that arise at the lowest energy scale),
\ba
\int \d^4x {\mathcal L}_{\rm int}&\supset& \int \d^4x \(\frac{\pi_0'(r)}{r \Lambda^6} \phi (\p_r^2 \phi)(\p_\Omega^2 \phi)+
\(\frac{1}{\Lambda^3}+\frac{\pi_0''(r)}{\Lambda^6}\)(\p_\Omega^2 \phi)^2 +\cdots\)\\
%&\supset&\int \d^4 \hat x \(\frac{\pi_0'(r)}{r \Lambda^6}\frac{1}{\(\e Z_\Earth\)^{3/2}} \hat \phi (\hat \p_r^2 \hat \phi)(\hat \p_\Omega^2 \hat \phi)+
%\(\frac{1}{\Lambda^3}+\frac{\pi_0''(r)}{\Lambda^6}\)\frac{1}{\e \(\e Z_\Earth\)^{3/2}}(\hat \p_\Omega^2 \hat \phi)^2 +\cdots\)\nn \\
&\supset&\int \d^4 \hat x \(\frac{1}{\hat{\Lambda}_{\Earth,1}^3} \hat \phi (\hat \p_r^2 \hat \phi)(\hat \p_\Omega^2 \hat \phi)+
\frac{1}{\hat{\Lambda}_{\Earth,2}^{3}}(\hat \p_\Omega^2 \hat \phi)^2 +\cdots\)\,,
\ea
with the redressed interaction scales,
\ba
\hat{\Lambda}_{\Earth,1}&=&\Lambda^2 \(\frac{R_\Earth}{\pi'_0}\)^{1/3}\(\e Z_\Earth\)^{1/2}\\
\hat{\Lambda}_{\Earth,2}&=&\Lambda^2 \(\frac{\e}{\pi''_0}\)^{1/3}\(\e Z_\Earth\)^{1/2}\sim \Lambda {\e}^{1/3}\(\e Z_\Earth\)^{1/2}\,.
\ea
Now going back into the original coordinates, if we read the redressed scale as a scale in the orthoradial direction then,
\ba
{\Lambda}_{\Earth,1,2}=\frac{1}{\sqrt{\e}} \hat{\Lambda}_{\Earth,1,2}
\ea
This leads to the following redressed interaction scales
\ba
{\Lambda}_{\Earth,1}&\sim & \Lambda Z_\Earth^{1/3} \\
{\Lambda}_{\Earth,2}&=&\Lambda \e^{1/3}\sqrt{Z_\Earth}\,.
\ea
Being extremely conservative and taking the na\"ive value for $\e$ as given in \eqref{naive epsilon}, we get
\ba
{\Lambda}_{\Earth,1}&\sim & 5\times 10^6 \Lambda \sim (20 {\rm cm})^{-1} \\
{\Lambda}_{\Earth,2}&\sim& 3\times 10^5 \Lambda \sim (30 {\rm cm})^{-1} \,,
\ea
assuming $\Lambda^3 \sim (H_0^2 \mpl)$.
This differs from the result of $(0.4 {\rm km})^{-1}$ obtained in \cite{1211.6001} by about three orders of magnitude. The reason for the discrepancy lies in the fact that the operator $(\p_\Omega^2 \phi)^2$ is not enhanced by the coefficients $\frac{\pi_0'(r)}{r \Lambda^6}$ as is the case for the other operators  of that form.  This is due to the specific structure of the Galileon interactions. As we show here, this operator has a coefficient going as $\pi_0''(r)$ rather than $\frac{\pi_0'(r)}{r}$. Since the hierarchy between $Z_{\Omega \Omega}$ and $Z_{tt}$ was precisely coming from the fact that around spherically symmetric configuration $\pi_0''(r)\ll \frac{\pi_0'(r)}{r}$ and so that operator comes in with a much larger energy scale than one could have anticipated at first sight\footnote{When dealing with the mixing $h^{\mu\nu}X^{3}\mn$, a confusion of similar nature was made in \cite{1211.6001}. Indeed, thanks to the very precise nature of the interactions in massive gravity, operators of the form $(\p^2 \pi)^3$ always appear in such a combination so that they form a total derivative, or in other words, there are no operators of the form $(\p^2 \pi)^3$ in massive gravity, which is the essence of its ghost-free construction \cite{dRG,dRGT}. Instead when considering an operator of the form $h (\p^2 \pi)^3$, its relevant contribution arises after integrations by parts to give rise to the operator $(\p^2 h)(\p \pi)^2 (\p^2 \pi)$, which is of course similar to $^*R^{\mu\nu\alpha\beta}\p_\mu \pi \p_\alpha \pi D_\nu D_\beta \pi$, this is an operator of dimension 7 rather than 9 which becomes relevant at a larger energy scale. However more importantly, as mentioned previously, there are no stable solutions which exhibit the Vainshtein mechanism when this mixing term is present.}.

Being even more conservative and translating instead the redressed scales  as distance scales along the radial direction, we would then have instead ${\Lambda}_{\Earth,1,2}\sim \hat{\Lambda}_{\Earth,1,2}$, leading to a smaller scale (or larger distance scale along the radial direction). Then for a more realistic value of $\e \lesssim 10^{-5}$, we would get instead ${\Lambda}_{\Earth,1}\sim (70{\rm m})^{-1}$, for $\Lambda\sim (H_0^2\mpl)^{1/3}$ as a radial distance scale (and $\Lambda_{\Earth, 2}\sim (1{\rm m})^{-1}$). However any realistic kind of matter located around the experiment itself would completely wash out this hierarchy and set $\e$ close to $1$, giving back ${\Lambda}_{\Earth,1,2}\sim {\rm few}\times (10 {\rm cm})^{-1}$ for $\Lambda\sim (H_0^2\mpl)^{1/3}$.
Since any test of Newton's law requires an apparatus which itself is of the order of a fraction of a metre, and is itself relatively massive, we expect the mass present at these distance scales to redress the strong coupling scale further. More work needs to be done to fully explore the influence of other environmental factors to the redressing of this scale.

\section{The Bigger Picture}
\label{sec:BiggerPicture}

As we have seen in the previous sections, massive gravity and generic Galileons have a relatively low energy scale at which perturbation theory breaks down. However, it is not because perturbation theory breaks down that we suddenly expect large corrections to Newton's law. If anything, we would expect the Vainshtein mechanism to work even better in this non-perturbative regime and to decouple the field even more. The scales $\Lambda_3$, $\Lambda_{\Earth}$ or $\Lambda_{\rm redressed}$ may or may not be interpreted as a cutoff in a usual sense, \ie it is not necessarily the case that new degrees of freedom come in at this scale. One may imagine it is the scale of strong coupling, analogous to $\Lambda_{\rm QCD}$, and non-perturbative tools must be developed to understand what happens at these scales.
For this reason it is important to stress that the scale at which perturbation theory breaks down cannot be used to put a lower bound on the graviton mass as has been argued in \cite{1211.6001}.

Interestingly for massive gravity the stability analysis of this paper constrains us to the restricted Galileons, considered in \cite{BCG}, for which $\beta=0$. In this case the only scale entering into the decoupling limit is $\Lambda=\Lambda_3/\alpha^{1/3}$. Since both $\alpha$ and $\beta$ are free parameters which satisfy a non-renormalization theorem \cite{dRGHP2}, the scale of strong coupling is completely independent of the mass of the graviton. Furthermore since essentially all observational constraints so far, constrain only the scale $\Lambda=\Lambda_3/\alpha^{1/3}$ \footnote{For instance in the calculation of the earth-moon orbit for lunar laser ranging \cite{LLR}, or pulsar radiation \cite{deRham:2012fw}, the decoupling limit description is sufficient to calculate the fifth forces or scalar radiation, and is hence determined only by the scale $\Lambda=\Lambda_3/\alpha^{1/3}$.} and not $\Lambda_3$ directly, they do not impose any constraints directly on the graviton mass. The later must be constrained by cosmology where the decoupling limit is not appropriate, rather than solar system/astrophysical gravitational tests \cite{LLR,deRham:2012fw}.

The results of this and the previous paper \cite{BCG} have important implications for the discussion on the possible existence of superluminalities, and UV completion. For instance, as argued in \cite{Adams}, generic Galileons violate the conditions for analyticity of the S-matrix in Minkowski space-time. However, the very definition of an S-matrix assumes the switching off of interactions at infinity consistent with the assumptions of asymptotic flatness. What we have seen is that introducing a single source into the theory is in conflict with asymptotically flat boundary conditions, at least in the decoupling limit, for stability reasons. As such the Minkowski S-matrix for the decoupling limit theory is not an appropriate description, and the existence or not of its analyticity is a mute point. This is not to say that these theories do not need to have a fundamentally different non-perturbative/UV completion than traditional Wilsonian effective field theories, but rather that their failure to satisfy usual analyticity properties in the decoupling limit by no means precludes the existence of a non-perturbative completion, even one for which the physics is potentially fundamentally (sub-)luminal.

\section*{Acknowledgements}

We would like to thank Matteo Fasiello, Lavinia Heisenberg, Andrew Matas and David Pirtskhalava for useful comments on the manuscript.
LB is supported by funds provided by the University
of Pennsylvania, GG is supported by NSF grant PHY-0758032 and NASA grant
NNX12AF86G S06, AJT was supported in part by the Department of Energy under grant DE-FG02-12ER41810.

\renewcommand{\theequation}{A-\Roman{equation}}
\setcounter{equation}{0}

\section*{Appendix A}

In order to find the most general ansatz for $\pi$, one has to satisfy the following conditions:

(i) Spherical symmetry requires $\pi(r,t)$ to be a function of $r$ and $t$ only.

(ii) For the configuration to be static the effective energy-momentum tensor ($T_{\mu\nu}^{\rm eff}\equiv G_{\mu\nu}$) must satisfy the following
\beq
T_{0i}^{\rm eff}=0 \quad \text{and} \quad \partial_t T_{00}^{\rm eff}=0.
\eeq
Here, the first expression requires the vanishing momentum density while the latter one assures time-independence of the energy density. One could notice that because of the energy-momentum conservation (Bianchi's identity), these two constraints are not independent from each other. However, nothing prohibits  requiring both of them individually.
Notice, that these conditions are the direct consequence of \eqref{anz}.

It follows from the Einstein's equation, that the condition of vanishing momentum-density reads as follows
\beq
&&\left[ \left(- 1+\frac{\alpha}{\Lambda^3}\Pi_{kk}+\frac{3 \beta}{\Lambda^6}(\Pi_{kk}^2-\Pi_{kl}\Pi_{kl})  \right)\delta_{ij} \right. \nonumber \\
&& \left.-\frac{\alpha}{\Lambda^3} \Pi_{ij}+ \frac{6 \beta}{\Lambda^6}(\Pi_{jk}\Pi_{ki}-\Pi_{ij}\Pi_{kk})\right]\Pi_{0j}=0.
\label{zero0j}
\eeq
For simplicity let us imagine that $\Pi_{0i}$ and $\Pi_{ij}$ are independent variables. Since, in that case we have a linear system of  algebraic equations at hand, with $\Pi_{0j}$ unknown. There exists a non-trivial solution to \eqref{zero0j} if and only if the determinant of the matrix in brackets vanishes. Otherwise, we end up with $\partial_0 \partial_r \pi=0$, which is consistent with our final ansatz \eqref{fin}. With the assumption of spherical symmetry
\beq
\Pi_{ij}=\frac{\pi^\prime}{r}\delta_{ij}+n_i n_j(\pi^{\prime\prime}-\frac{\pi^\prime}{r}),
\eeq
the above-mentioned condition becomes
\beq
\left(-1+\frac{\alpha}{\Lambda^3}\left(\pi^{\prime\prime}+\frac{\pi^\prime}{r}\right)+\frac{6\beta}{\Lambda^6}\pi^{\prime\prime}\frac{\pi^\prime}{r}\right)^2\left(-1-\frac{2\alpha}{\Lambda^3}\frac{\pi^\prime}{r}+\frac{6\beta}{\Lambda^6}\left(\frac{\pi^\prime}{r}\right)^2\right)=0.
\label{detsol}
\eeq
From (\ref{detsol}) and the time independence of $T^{\rm eff}_{0\mu}$ follows that the most general ansatz relevant to us is
 \beq
\pi=T(t)+\pi_0(r), \qquad \forall ~T,\pi_0.
\label{fin}
\eeq
If we further require the time-independence of the effective stress tensor $T^{\rm eff}_{ij}$, we arrive at the following ansatz
\beq
\pi=\frac{c}{2}\Lambda^3t^2+\pi_0(r),
\eeq
where we could add terms constant and linear in time, however those terms are irrelevant because of the Galilean symmetry.

\end {document}